\documentclass[11pt]{amsart}

\usepackage[textwidth=0.75in,textsize=small]{todonotes}
\usepackage{amsthm}
\usepackage[all,pdf,cmtip]{xy}
\usepackage{amssymb,amsfonts,amsmath}
\usepackage{amscd}
\usepackage[english]{babel}
\usepackage{import}
\usepackage{mathtools}
\usepackage[colorlinks=true]{hyperref}
\usepackage{subfiles}
\usepackage{amsmath,amssymb,amscd,amsfonts,graphicx,latexsym,epsfig,psfrag,url,color,xr,mathtools,enumitem,transparent,fullpage,hyperref,float,mathrsfs,wrapfig,rotating,array}
\usepackage[foot]{amsaddr}
\usepackage{enumitem}

\newtheorem{theorem}{Theorem}[section]

\theoremstyle{definition}

\theoremstyle{remark}

\theoremstyle{plain}
\newcommand{\thistheoremname}{}
\newtheorem{genericthm}[theorem]{\thistheoremname}

\newtheorem*{genericthm*}{\thistheoremname}
\newenvironment{namedthm*}[1]
  {\renewcommand{\thistheoremname}{#1}%
   \begin{genericthm*}}
  {\end{genericthm*}}
  
\newcommand\qu{/\kern-.7ex/} 
\newcommand\lqu{\backslash \kern-.7ex \backslash}

\newcommand{\wh}{\widehat}

\newcounter{qcounter}

\newcommand\quotient[2]{
        \mathchoice
            {% \displaystyle
                \text{\raise1ex\hbox{$#1$}\Big/\lower1ex\hbox{$#2$}}%
            }
            {% \textstyle
                #1\,/\,#2
            }
            {% \scriptstyle
                #1\,/\,#2
            }
            {% \scriptscriptstyle
                #1\,/\,#2
            }
    }

\newcommand\quoti[2]{
                \text{\raise1ex\hbox{$#1$}/\lower1ex\hbox{$\scriptstyle#2$}}
  }

\newcommand\quot[2]{
                \text{\raise1ex\hbox{$#1\!\!$}/\lower1ex\hbox{$\!\scriptstyle#2$}}
  }

\newcommand\quo[2]{
                \text{\raise.8ex\hbox{$\scriptstyle#1\!$}/\lower.8ex\hbox{$\!\scriptstyle#2$}}
  }

\newcommand\qq[2]{
                \text{\raise.8ex\hbox{$#1\!$}/\lower.8ex\hbox{$#2$}}
}

\begin{document}

\title{A comparison of group testing architectures for COVID-19 testing}

\author{J.\ Batson}
\email{\href{mailto:joshua.batson@czbiohub.org}{joshua.batson@czbiohub.org}}

\author{N.\ Bottman}
\email{\href{mailto:bottman@usc.edu}{bottman@usc.edu}}

\author{Y.\ Cooper}
\email{\href{mailto:yaim@math.ias.edu}{yaim@math.ias.edu}}

\author{F.\ Janda}
\email{\href{mailto:janda@math.ias.edu}{janda@math.ias.edu}}
\date{\today}

\maketitle

\begin{abstract}

An important component of every country's COVID-19 response is fast and efficient testing --- to identify and isolate cases, as well as for early detection of local hotspots.
For many countries, producing a sufficient number of tests has been a serious limiting factor in their efforts to control COVID-19 infections.
\emph{Group testing} is a well-established mathematical tool, which can provide a substantial and inexpensive expansion of testing capacity.
In this note, we compare several popular group testing schemes in the context of qPCR testing for COVID-19.
%We include example calculations, where we indicate which testing architectures yield the greatest efficiency gains in various settings.
We find that in practical settings, for identification of individuals with COVID-19, Dorfman testing is the best choice at prevalences up to 30\%, while for estimation of COVID-19 prevalence rates in the total population, Gibbs--Gower testing is the best choice at prevalences up to 30\% given a fixed and relatively small number of tests. For instance, at a prevalence of up to 2\%, Dorfman testing gives an efficiency gain of 3.5--8; at 1\% prevalence, Gibbs--Gower testing gives an efficiency gain of 18, even when capping the pool size at a feasible number.
This note is intended as a helpful handbook for labs implementing group testing methods.

%An important component of every country's COVID-19 response is fast and efficient testing --- to identify and isolate cases, as well as for early detection of local hotspots.
%For many countries, producing a sufficient number of tests has been a serious limiting factor in their efforts to control COVID-19 infections.
%\emph{Group testing} is a well-established mathematical tool, which can provide a substantial and inexpensive expansion of testing capacity.
%In this note, we compare several well-established group testing schemes in the context of qPCR testing for COVID-19.
%We include example calculations, where we indicate which testing architectures yield the greatest efficiency gains in various settings.
%We find that for identification of individuals with COVID-19, Dorfman testing is usually the best choice, while for estimation of COVID-19 prevalence rates in the total population, Gibbs--Gower testing usually provides the most accurate estimates given a fixed and relatively small number of tests. 
%This note is intended as a helpful handbook for labs implementing group testing methods.
\end{abstract}

\tableofcontents

\section{Introduction}

Group testing is a standard technique for testing a population for a disease with a low incidence rate, while using fewer tests than if every individual were tested.
This schema goes back to at least 1943, when Dorfman \cite{dorfman} proposed screening prospective soldiers for syphilis by using two-stage hierarchical group testing.
Over the ensuing decades, the theory of group testing has been extensively developed, as surveyed in \cite{hughes-oliver}.
Group testing is a powerful tool, but not a silver bullet, and it has limitations.
In this note, we highlight some COVID-19 testing situations where group testing can give the biggest gains.
We also provide some guidelines, and note some pitfalls to watch out for.

There are two basic applications: the \emph{classification problem}, in which we aim to classify every individual as positive or negative, and the \emph{estimation problem}, in which we aim to estimate the prevalence of the disease in the population.
We will consider both problems in this note.

A basic example of using group testing for classification might proceed like this: instead of using 15 tests for 15 people, use 10 tests for 10 pooled groups of 5 individual samples each.
If, on average, only one of those 10 groups is positive, the remaining 5 tests could be used to test the individual samples in that positive group.
Therefore, in this example, 15 tests could reach 50 people on average, instead of only 15.
In general, when the incidence rate of the disease is low and the test is accurate, this method can provide accurate results while using many fewer tests.

If the number $T$ of tests at our disposal is fixed, the number of individuals that we can test depends on the incidence rate.
Several groups have worked out the optimal pooling architecture in various settings.
One notable example is the \texttt{R} package \texttt{binGroup} \cite{bingroup}, which can systematically deduce the optimal testing architecture.

%In this note, we provide some sample calculations, using current estimates for coronavirus infection rates, to illustrate how pooled testing can help, both at the individual level and the population level --- that is, both to tell whether a particular person has coronavirus, and to monitor the overall prevalence rate in the population.

This note is intended as a helpful guide for labs implementing group testing methods.  To make this note accessible to the broadest audience, we include several background sections that not every reader will need.  
The authors are happy to help any lab interested in leveraging these techniques to design optimal architectures given local constraints.

Here is a brief overview of this note:

\bigskip

\noindent
{\bf \S\ref{s:executive_summary}:}
We provide a concise executive summary, which decision-makers should consult for headline-level recommendations about the efficacy of group testing in various settings.

\medskip

\noindent
{\bf \S\ref{s:biology}:}
We discuss COVID-19 testing both for active infections and for antibodies.  We also discuss relevant group testing architectures for each --- Dorfman testing, array testing, hypercube testing for qPCR tests and Gibbs--Gower testing for antigen tests, with an eye toward logistical considerations.

\medskip

\noindent
{\bf \S\ref{s:math}:}
First, we outline several different goals of group testing.
Then, we describe and compare those group testing architectures that we believe are most applicable for COVID-19 testing.

\medskip

\noindent
{\bf \S\ref{s:not_helpful}:}
While group testing can be a powerful tool, there are several aspects --- practical and theoretical --- that one must regard carefully in applications.
We summarize these potential pitfalls.

\medskip

\noindent
{\bf \S\ref{s:addl_architectures}:}
Besides those described in \S\ref{s:math}, there are several more group testing architectures in the literature.
We summarize these, and explain why they are likely not to be useful for COVID-19 testing. 

\medskip

\noindent
{\bf \S\ref{s:guidelines}:}
We compare the efficiency gains produced by the architectures we described in \S\ref{s:math}, in the context of both classification and estimation.
This can be thought of as the basic supporting material for \S\ref{s:executive_summary}.

\medskip

\noindent
{\bf \S\ref{s:examples}:}
We work out in some detail several extended examples of applying the architectures in \S\ref{s:math} to COVID-19 testing.

\medskip

\noindent
{\bf \S\ref{s:conclusion}:}
We draw several fundamental conclusions.

\medskip

\noindent
{\bf \S\ref{s:formulas}:}
In this appendix, we provide the supporting calculations for the body of the paper.

\medskip

\section{Executive summary}
\label{s:executive_summary}

\medskip

In settings where tests are expensive or limited, group testing is a mathematical strategy that can substantially expand the number of people tested with a fixed number of tests.
Testing for COVID-19 is a good candidate for using group testing strategies to expand testing capacity worldwide.

\begin{figure}[H]
\centering
\def\svgwidth{0.8\columnwidth}
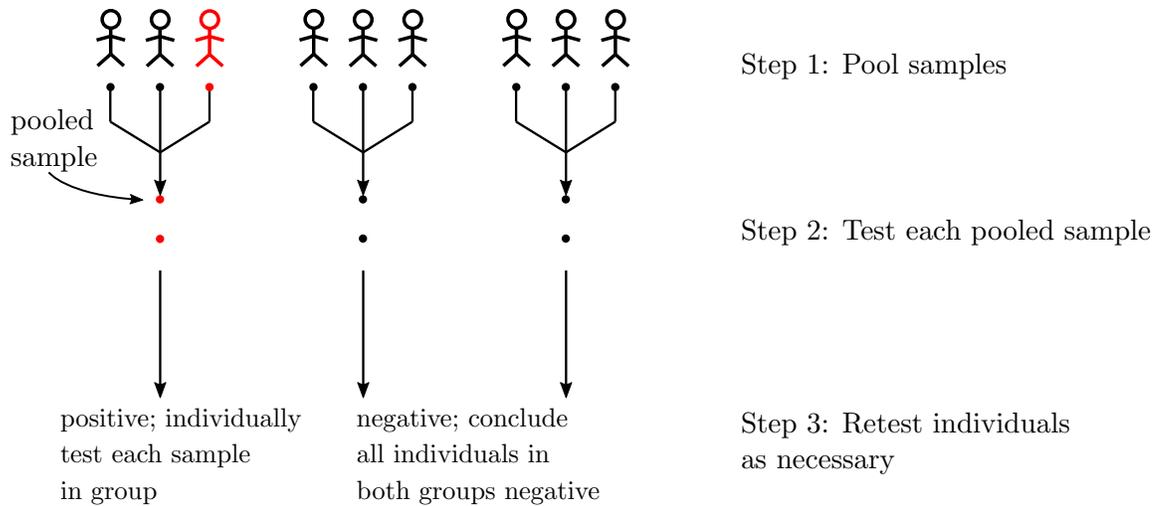
\bigskip\bigskip
\caption{
A schematic of Dorfman testing, a group testing scheme in which pools of samples are tested, and then individual samples from positive pools are retested to identify positive individuals.
Here, a pool size of 3 is used to find the positive individual in a group of 9 using only 6 tests.
%A schematic of Dorfman testing, applied here with pool size 3.
}
\end{figure}

There are two primary methods of testing for COVID-19:
\begin{itemize}
\item
For detecting active infections, qPCR is used to detect viral RNA in the host.

\item
For detecting past infections, antibody tests are used to detect immune response to COVID-19 virus.
\end{itemize}
%{\bf Based on the expected prevalence rates as well as the details of the testing protocols for these two types of tests, group testing is likely to provide substantial benefits for qPCR testing for COVID-19, while it will have more limited application for antibody testing for COVID-19.}
Based on the expected prevalence rates as well as the details of the testing protocols for these two types of tests:
\begin{itemize}
\item
{\bf Group testing for viral RNA with qPCR can provide substantial benefits} with practical pool sizes between 4 and 24.

\item
{\bf Group testing for antibodies may be less effective,} because group testing is less effective at higher expected prevalences.
In a region with expected positivity rate sufficiently low, our recommendations for qPCR group testing continue to hold for antibody testing.
\end{itemize}

For labs implementing group testing for qPCR testing, the best practice is to have a
%N mathematician
mathematician, statistician, or data scientist
on staff to regularly monitor and re-optimize the group testing architecture.  This person should watch especially for dilution effects, changes in prevalence rate, and in the case of estimation, bias in the estimator.  

In a test-limited setting, there are multiple testing goals one might try to optimize with a limited number of tests: classification, estimation, verified positives, and verified negatives, as described in \S\ref{s:math}.
{\bf In settings that make it impossible to have a dedicated
%N mathematician
mathematician, statistician, or data scientist on staff,
here are reasonably good one-size-fits-all strategies for classification and estimation.}

\begin{figure}[H]
\centering
\def\svgwidth{0.8\columnwidth}
%% Creator: Inkscape 1.0.1 (c497b03c, 2020-09-10), www.inkscape.org
%% Accompanies image file 'schematic_2.pdf' (pdf, eps, ps)
\begingroup%
  \makeatletter%
  \providecommand\color[2][]{%
    \errmessage{(Inkscape) Color is used for the text in Inkscape, but the package 'color.sty' is not loaded}%
    \renewcommand\color[2][]{}%
  }%
  \providecommand\transparent[1]{%
    \errmessage{(Inkscape) Transparency is used (non-zero) for the text in Inkscape, but the package 'transparent.sty' is not loaded}%
    \renewcommand\transparent[1]{}%
  }%
  \providecommand\rotatebox[2]{#2}%
  \newcommand*\fsize{\dimexpr\f@size pt\relax}%
  \newcommand*\lineheight[1]{\fontsize{\fsize}{#1\fsize}\selectfont}%
  \ifx\svgwidth\undefined%
    \setlength{\unitlength}{360.98533025bp}%
    \ifx\svgscale\undefined%
      \relax%
    \else%
      \setlength{\unitlength}{\unitlength * \real{\svgscale}}%
    \fi%
  \else%
    \setlength{\unitlength}{\svgwidth}%
  \fi%
  \global\let\svgwidth\undefined%
  \global\let\svgscale\undefined%
  \makeatother%
  \begin{picture}(1,0.47746614)%
    \lineheight{1}%
    \setlength\tabcolsep{0pt}%
    \put(0,0){\includegraphics[width=\unitlength,page=1]{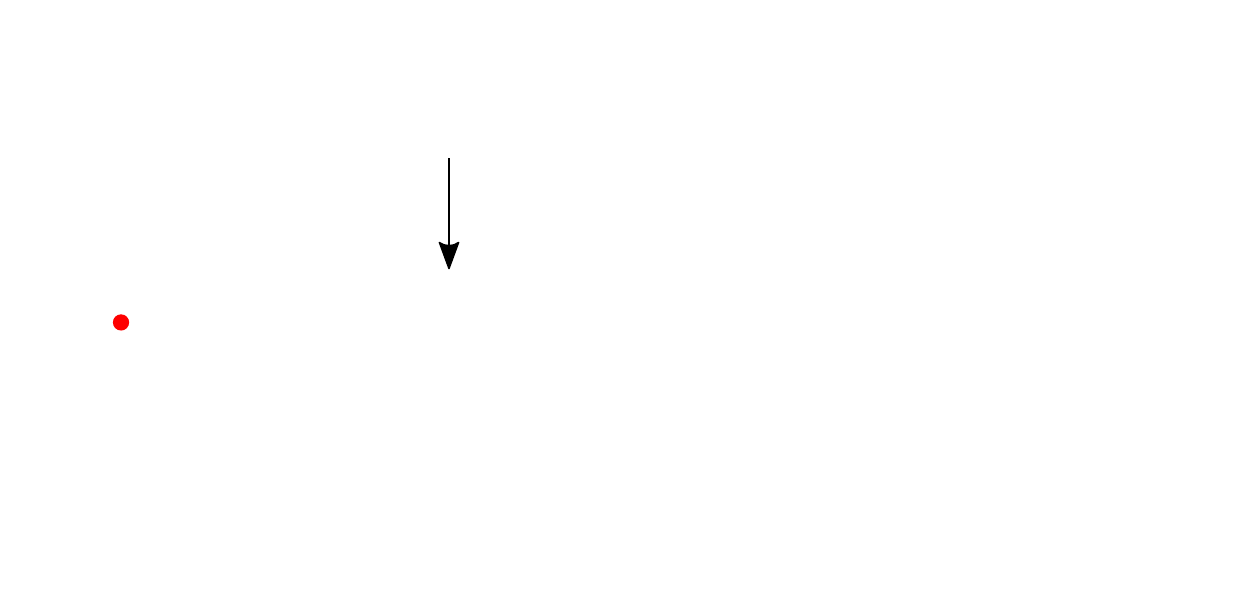}}%
    \put(0.79249909,0.45954996){\makebox(0,0)[lt]{\lineheight{1.25}\smash{\begin{tabular}[t]{l}Step 1: Pool samples\\taken from larger\\population\end{tabular}}}}%
    \put(0,0){\includegraphics[width=\unitlength,page=2]{schematic_2.pdf}}%
    \put(0.79249909,0.20010846){\makebox(0,0)[lt]{\lineheight{1.25}\smash{\begin{tabular}[t]{l}Step 2: Test each\\pooled sample\end{tabular}}}}%
    \put(0.79249909,0.03656797){\makebox(0,0)[lt]{\lineheight{1.25}\smash{\begin{tabular}[t]{l}Step 3: Estimate\\prevalence\end{tabular}}}}%
    \put(0.11503786,0.03241267){\makebox(0,0)[lt]{\lineheight{1.25}\smash{\begin{tabular}[t]{l}estimated prevalence $\wh p = 1 - \left(1-\frac{t_+}t\right)^{1/b}$\end{tabular}}}}%
  \end{picture}%
\endgroup%

\bigskip\bigskip
\caption{
A schematic of Gibbs-Gower testing, a group testing scheme in which pools of samples are tested to estimate the prevalence of the disease in a population.
A formula relates the number of positive pools to the prevalence: for the above example with a pool size of 7, the estimated prevalence of $\hat{p} = 1 - \left(1-\frac{t_+}{t}\right)^{\frac{1}{b}} \approx 0.056$, while the true prevalence in this population was 0.071.
For simplicity, in this example the whole population was tested, though Gibbs-Gower testing is usually applied to a subsample of the population.
%A schematic of Gibbs--Gower testing, applied here with pool size 7.
%In this example, the method provides an estimated prevalence of $\hat{p} = 1 - \left(1-\frac{t_+}{t}\right)^{\frac{1}{b}} \approx 0.056$, while the true prevalence in this population was 0.071.
%For simplicity, in this example the whole population was tested, though Gibbs-Gower testing is usually applied to a subsample of the population.
}
\end{figure}

\begin{itemize}
\item
For classification: use {\bf simple Dorfman testing (\S\ref{ss:dorfman}), with pool size 3--8, for prevalence up to 30\%.}
Use the following chart, which shows prevalence, suggested pool size, and efficiency gain.

\medskip

\begin{center}
\begin{tabular}[center]{|c|c|c|}
  \hline
  prevalence & optimum pool size & efficiency gain
  \\
  \hline
  \hline
  12.5--30\% & 3 & 1--1.5
  \\
  \hline
  6.6--12.5\% & 4 & 1.5--2
  \\
  \hline
  4.1--6\% & 5 & 2--2.5
  \\
  \hline
  2.8--4.1\% & 6 & 2.5--3
  \\
  \hline
  2--2.8\% & 7 & 3--3.5
  \\
  \hline
  up to 2\% & 8 & 3.5--8
  \\
  \hline
\end{tabular}
\end{center}

\medskip

\item
For estimation: use {\bf Gibbs--Gower testing (\S\ref{ss:gibbs-gower}).}
Use \eqref{eq:GG_estimator} to calculate an estimation of the true prevalence.
Use the pool size specified in the following chart.
This chart shows the optimum pool size (capped at 20) for minimizing the number of test for a given prevalence rate necessary for bringing the relative error of the estimate below $15\%$, as well as the efficiency gain at the optimum pool size compared to the number of individual tests necessary to achieve the same relative error.

\medskip

\begin{center}
\begin{tabular}[center]{|c|c|c|}
  \hline
  prevalence & optimum pool size ($\leq 20$) & efficiency gain
  \\
  \hline
  \hline
  0.1\% & 20 & 20
  \\
  \hline
  0.2\% & 20 & 20
  \\
  \hline
  0.5\% & 20 & 19
  \\
  \hline
  1\% & 20 & 18
  \\
  \hline
  2\% & 20 & 16
  \\
  \hline
  5\% & 20 & 11
  \\
  \hline
  10\% & 13 & 5.8
  \\
  \hline
  20\% & 6 & 2.9
  \\
  \hline
  30\% & 4 & 2.0
  \\
  \hline
\end{tabular}
\end{center}

\end{itemize}

\section{Biochemical considerations}
\label{s:biology}

We begin with a brief overview of some standard testing procedures for COVID-19.  

\subsection{Testing for active infection: qPCR}

In the context of COVID-19, qPCR tests are the gold standard for detecting active infections.  A typical testing protocol includes the following steps.

\begin{enumerate}
\item
A sample is collected from the individual being tested.
Common methods of sampling are a nasalpharangeal swab (NP), oralpharangeal swab (OP), anterior nasal swab (AN), or saliva.

\item
Samples arrive at a lab as a swab in media.
Transport media include RNA shield, Viral Transport Media (VTM), Phosphate-buffered Saline (PBS), and pure Saline.

\item
Samples are inactivated (except if the sample came in RNA shield, in which case it is already inactivated).

\item
Swabs are extracted and discarded.  
\end{enumerate}

\subsubsection{Research lab}

In a research lab setting, typical next steps are the following.

\begin{enumerate}[resume]
\item
Pipette aliquots from each sample into a 96-well plate.  Depending on the lab, this is done manually, or with a liquid handler such as a Hamilton Starlet.

\item
Then, the remaining steps are done in parallel on the plates.  The reagents for each of the following steps are typically handled by a liquid handling robot, such as an Agilent Bravo or a Hamilton Starlet.

\item
RNA extraction.

\item
RT-PCR (turning RNA into cDNA).

\item
qPCR.
\end{enumerate}

\subsubsection{Private lab}

In the United States, currently a large proportion of the qPCR tests are being processed at private labs such as LabCorp and Quest Diagnostics.  

In these settings, it is typical after (4) for the process to be handled by high-throughput sample-to-answer machines, such as the Cobas 6800 from Roche or the M2000 from Abbott.
In this case, the sample tubes are racked and loaded, and all subsequent steps are performed by the machine without human intervention.
One final result is returned per well. 

\subsubsection{On-site testing}

Another important direction is the development of rapid test machines that can screen people on-site.
This is likely to be important in many settings, including airports and arenas.

An example of such technology is the Abbott ID Now system.
In this system, a sample is taken from an individual, inserted into the machine, which performs rapid qPCR on the single sample and returns a positive or negative result.   

\subsubsection{Low-resource settings}

Thus far, we have described qPCR settings at well funded public and private labs in the US.
In many communities in the US and around the world, COVID-19 testing is and will be done in more resource constrained settings.
In such settings, the benefits of group testing will be even greater.  

In areas with fewer resources, there may be shortages of reagents and machines for performing qPCR.
Furthermore, labor costs are likely to be lower, which would make it relatively cheaper to implement group testing schemes.

\subsubsection{Group testing for qPCR testing}
In many settings, qPCR is a good candidate for group testing.
Group testing cannot provide efficiency gains at all stages of the testing protocol, but can be applied at the lab stage, for the performance of the qPCR measurement.
In research labs, group testing could be integrated into the testing protocol at many labs by a change in protocol, possibly in conjunction with modifications in software for liquid-handling robots.
In private labs using sample-to-answer machines, the initial cost for implementing group testing would be higher as it would require modification of the software in a complex machine, but the savings could be substantial.
Pooled testing could make sense for on-site testing where low prevalences are expected.
In that setting, it could give substantial cost as well as time savings for the average subject.
Finally, in low-resource settings, group testing is likely to have the greatest benefit, as the cost to group testing is primarily in labor, which would be relatively less expensive, and the savings would be primarily in reagents and machines for running qPCR, which would be relatively more constrained.  

\subsection{Testing for prior infection: Antibody testing}

Going forward, it will be important both to detect active COVID-19 infections, as well as past infections.
The purpose of the qPCR tests discussed above are to detect active infections.
Concurrently, antibody tests have been developed to detect IgG and IgM antibodies to COVID-19, which present in most patients who have had a COVID-19 infection between 1 and 2 weeks after the initial infection.
It is not currently known how long antibodies continue to circulate in the blood after an infection.

For antibody testing, a blood sample is drawn from the subject.
How the sample is then tested for antibodies to COVID-19 depends on the test setting.

\subsubsection{Research lab}

In a research lab, typically samples will be loaded onto 96-well plates.
This will be followed by an Enzyme-linked immunosorbent assays, or ELISA, assay to detect antibodies.

\subsubsection{Private lab}

In a private lab, typically samples are processed by a sample-to-answer machine, for example the Abbott ARCHITECT.
These machines have high specificity and are able to process high volumes, such as several thousand tests per day.

\subsubsection{On-site testing}

Antibody testing may also be done in a point-of-care or onsite setting.
In this setting, lateral flow assays (LFAs) and immunochromatographic strip tests (ICTs) are common.
These often resemble home pregnancy tests --- the sample flows up a strip and the presence of antibodies is signaled by a change in color of the test region.

These tests are both less specific and less sensitive than other methods, and are not a good candidate for pooling.

\subsubsection{Low-resource settings}
In low resource settings, the most likely antibody tests are LFAs, as they are relatively inexpensive and easy to deploy.

\subsubsection{Group testing for antibody testing}

The most significant factor in determining the efficiency gain from group testing is the prevalence rate in the tested population.
Even in the theoretical limit, there is little or no benefit to using group testing once the prevalence rate exceeds 30\%.  

Estimates for the current prevalence rate for COVID-19 antibodies in the general population in metro areas of the US currently range from a few percent to as much as 30\%.
As the pandemic evolves, these numbers will only go up.
Therefore, group testing will provide limited benefits in the setting of antibody testing.  

However, it is possible that in limited settings, it is worth applying group testing to antibody testing for COVID-19, especially early on.

\section{Several basic group testing architectures}
\label{s:math}

In this section, we will give a detailed overview of several important group testing architectures for both the classification and the estimation problems.
Before we proceed with this overview, we describe several goals for group testing.
%The most obvious goal is to identify all positive individuals while minimizing the number of tests.
%Group testing can be used for this purpose but there are also other possible testing objectives.
%They include the following:
\smallskip
\begin{itemize}
\item
{\bf Classification:} Classify each individual as being positive or negative.
This is the most classical application of group testing.
  
\medskip

\item
{\bf Estimation:}
Use test results to accurately estimate the prevalence in the population, without necessarily classifying each individual.

\medskip

\item
{\bf Verified negatives:}
With a given number of tests, maximize the number of individuals that can be certified (with high probability) to be negative.
  
\medskip
  
\item
{\bf Verified positives:}
With a given number of tests, maximize the number of individuals that can be certified (with high probability) to be positive.
\end{itemize}
\smallskip
The rationale behind the ``verified negatives'' goal is to clear individuals to return to work, for instance.
On the other hand, the rationale behind the ``verified positives'' goal might be to keep as many positive individuals at home as possible.
The ``verified positives'' testing goal is also known as ``total recall'' \cite{Broad}.

%{\it
%Overview of couple group testing schemes that we will recommend
%\begin{itemize}
%\item
%Dorfman testing.
%
%\item
%Hypercube testing.
%
%\item
%Gower--Gibbs prevalence.
%\end{itemize}
%}
%
%{\it
%Ideas:
%
%Keep campus safe:
%Do array testing, advice any in intersection of positive to stay at
%home, clear not in intersection
%
%Max positives:
%Do Sterrett, but don't test the rest of a batch after finding a positive
%
%Max negatives:
%Like Dorfman, but retest if positive. Certify negative batches.
%}

\subsection{Architectures for the classification problem}

\subsubsection{Dorfman testing (i.e.\ simple 1-layer group testing)}
\label{ss:dorfman}

Perhaps the simplest version of group testing is simple 1-layer group testing.
This was first suggested by Dorfman in 1943 as a method for screening military recruits for syphilis \cite{dorfman}, so we refer to this method as \emph{Dorfman testing}.

We begin with an overview of Dorfman testing.
Suppose that we have a population of $n$ individuals, and we would like to determine, using the minimal number of tests, whether each individual is positive or negative for a given disease.
We proceed as follows:
\begin{enumerate}
\item
Group the individuals into $n/b$ batches of $b$.

\item
For each batch, mix the samples into a single pooled sample.

\item
Using $n/b$ tests, test these pooled samples.

\item
For each pooled sample which tests negative, we deduce that all of the individuals in the corresponding batch are negative.
For each pooled sample which tests positive, we use $b$ tests to test each individual in the corresponding batch.
\end{enumerate}
When the incidence rate is low, Dorfman testing will use fewer tests than $n$, which is the number of tests that we would use if we follows a non-grouped approach and tested every individual.

\subsubsection{Array testing}

Next, we mention an alternative to the \emph{simple} or \emph{hierarchical} procedures described above: \emph{array testing}.
As described in \cite{PhatarfodSudbury}, and as the $d=2$ case in \cite[\S2]{hughes-oliver}, the population is divided into $n/b^2$ clusters of $b^2$ individuals.
Each cluster is arranged in a 2-dimensional array with $b$ rows and $b$ columns.
Each row is tested, and each columns is tested, using a total of $2b$ tests for each cluster.
If either the $i$-th row or the $j$-th column both test negative, then we can conclude that the $(i,j)$-th individual is negative.
There are two variants regarding how to treat the individuals whose row and column both test positive:
\begin{enumerate}
\item
Test each such individual, thus establishing precisely which individuals are positive.

\item
Presume that each such individual is positive, reducing the total number of tests used at the expense of a modest (so long as $\rho$ is small) rate of false positives.
\end{enumerate}
A feature of array testing is that the expected total number of tests (given below as \eqref{eq:num_tests_unbatched_array}) is a function of $b$ (as well as of $\rho$ and $n$) --- in contrast with Dorfman testing, where the expected number of tests is of the form $f(\rho)n$.
By choosing $b$ carefully, we can therefore optimize the number of individuals tested per test.

\subsubsection{Hypercube testing}

Finally, we mention a higher-dimensional version of array testing.
We are not aware of where this method was originally proposed, but it is referred to in \cite{hughes-oliver} as an established method, and goes back at least to 2000 \cite{BergerMandellSubrahmanya}; in the context of COVID-19 detection, it was the subject of the recent paper \cite{turok}.

We begin by choosing a \emph{dimension} $d \geq 2$, and dividing our population into $n/b^d$ clusters of $b^d$ individuals.
We arrange each cluster into a $d$-dimensional cube of side-length $b$.
(Of course, we do not perform this arrangement physically, since that would be impossible for $d \geq 4$.)
This serves to index our population by $d$ indices, so that each individual corresponds to a sequence $(i_1,\ldots,i_d)$ with $1 \leq i_j \leq b$ for each $j$.
For each cluster, we then use $db^{d-1}$ tests, which we use to do the analogue of testing the rows and columns --- that is, for each $j$ and choice of $i_1,\ldots,i_{j-1},i_{j+1},\ldots,i_d$, we test all the individuals with label of the form $(i_1,\ldots,i_{j-1},?,i_{j+1},\ldots,i_d)$.
We proceed analogously with array testing.

\subsection{Architectures for the estimation problem}

In the previous section, we focused on the goal of using group testing to \emph{classify} whether each individual in a population has a certain disease.
On the other hand, there are situations where we merely want to \emph{estimate the proportion} of infected individuals.

In this section, we focus on the latter problem.
Just as for classification, we will see that batching tests can be a valuable tool for the estimation problem.
In fact, group testing can be even more effective in this setting, because we can estimate the number of positive individuals in a positive group without further testing each individual in the group.

\subsubsection{Description of a simple group-testing algorithm for the estimation problem}
\label{ss:gibbs-gower}

We now describe the simplest way to apply group testing to the estimation problem.
This method was first proposed by Gibbs--Gower \cite{gibbs_gower} in 1960 (by the name
``multiple-transfer method'') \cite{gibbs_gower}, and further studied by Thompson in 1962 \cite{thompson}.
We refer to \cite[\S3.1.2]{hughes-oliver} for further review of the literature.

Suppose that we have a population of $n$ individuals, which we divide into $n/b$ groups of $b$ individuals.
Denote the true incidence rate by $p$.
The likelihood that a given group has at least one positive individual is then $P \coloneqq 1 - (1 - p)^b$.

Next, we test $t \leq n/b$ of the groups.
Denoting the number of groups testing positive by $t_+$, we obtain an estimate
\begin{equation*}
\hat P
\coloneqq
\frac{t_+}t
\end{equation*}
for $P$, and an estimate
\begin{align}
\hat p
&\coloneqq
1 - (1 - P)^{1/b}
\nonumber
\\
&=
1 - \left(1 - \frac{t_+}t\right)^{1/b}
\label{eq:GG_estimator}
\end{align}
for $p$.

\section{Points of caution}
\label{s:not_helpful}

There are many factors to consider in designing a group testing architecture for a given setting.  The utility of group testing and the optimal design depend very heavily on the details of the lab setting, test population, and so on.  Here we collect some common considerations.  

\subsection{Batch size}
 
The theoretically-optimal batch size for a group testing architecture can be very large.
For example, in \S\ref{s:examples} we consider the task of computing that to estimate prevalence with an RMSE of within $15\%$ of the true prevalence.
When we use Gibbs--Gower testing and assume that the true prevalence is $0.01\%$, the optimal batch size at $\rho = 0.3\%$ is 12150.
There are a number of practical reasons for why one would want to limit the batch size.

First, there are chemical and biological aspects of the test itself that must be considered.
For qPCR, there are studies such as \cite{kishony} that suggest that sensitivity remains high for groups as large as 32 or 64, and that it is possible to increase the number of cycles to counteract the effects of diluting samples.
However, one may reasonably want to limit the batch sizes to be comfortably lower than 32 in practice, to minimize the danger of errors introduced by dilution.

Secondly, every lab will have a limit on the average daily throughput.
One therefore wants to bound the total number of samples processed in any given run by the expected daily load of the lab.
For example, suppose a lab that processes 1000 samples a day is considering a 3-dimensional hypercube architecture.
They may desire both that the side length of the cube, i.e.\ the number of samples pooled for each chemical test, to be less than 16, as well as that the total number of samples in the cube be less than 1000.
In this case, the second constraint is stronger, and the lab should use cubes of side length less than or equal to 10.

\subsection{Variation in prevalence}

Another consideration is that the true prevalence in the sampled population may differ from the prevalence rate that the group testing architecture was optimized for.
This will decrease the efficiency gain of using group testing.  
In some regimes the efficiency gain is more sensitive to variation in true prevalence than others, and this is a factor that should be analyzed in any implementation of group testing.
A reasonable approach to this problem is to optimize the group testing architecture not for a specific prevalence but for a range of expected prevalences, to maximize the efficiency gain over a distribution of expected prevalences.  
Optimizing the batch size for the mean prevalence is usually a good approximation for this.

\subsection{Time and complexity considerations}

Dorfman's original testing architecture is now considered the standard basic setup for group testing.
There are many variations on Dorfman testing of increasing complexity, which provide even greater efficiency gains.
However, in a lab setting, there is a premium on using simpler architectures, which reduce the opportunities for introduced error.

Perhaps the most substantial consideration in the complexity of the architecture is the amount of time that the testing algorithm takes to run compared to non-pooled testing.
Architectures such as binary search or Sterrett testing require many sequential tests.
When each test can be run quickly, this may not be an issue.
However, for some tests --- such as qPCR, where each test takes several hours to run --- this may render certain group testing architectures infeasible.  

For this reason, in our general recommendations we restrict our attention to architectures that introduce only one or two additional rounds of testing.
However, when customizing a group testing architecture for a particular application, one may want to consider more complicated testing schemes.

\subsection{Bias in estimator}

For prevalence estimation using Gibbs--Gower testing, there is an additional factor which should be kept in mind: the computed estimator is biased.
This effect is especially pronounced when the prevalence increases.
For example, at a prevalence of 30\% and with a batch size of 30, the probability of a given batch testing positive is $\sim\!99.998\%$.
Unless one is testing an extremely large number of batches, it will therefore not be possible for this method to distinguish between a true prevalence of 100\% and a true prevalence of 30\%.
This is the underlying cause of the bias in the estimator.

Care should therefore be taken that the batch size is small enough to be sensitive to the largest expected prevalence.
This, of course, limits the efficiency gained by using Gibbs--Gower testing.
Despite this issue, Gibbs--Gower can provide a significant gain in many regimes.

\section{Monitoring dilution effects}
\label{s:dilution}

One of the most important factors to keep track of when implementing group testing is the introduction of false negatives due to dilution effects.
The viral load present in individuals infected with COVID-19 can vary over more than 9 orders of magnitude.
At viral loads at the low end of this spectrum, qPCR tests developed for COVID-19 detection can miss infected individuals.

Some ways in which individuals with low viral loads may go undetected is either that virus particles are not obtained in the initial sample (NP swab, AP swab, saliva, etc.), or that the virus is present in the initial sample but at such a low concentration that the average number of viral particles in the quantity of liquid pipetted into the well for performing qPCR is small.
At low concentrations, due to stochasticity, in a substantial percentage of such samples, no viral particles will be sampled from the initial sample.  

Pooled testing protocols where the amount of material taken from each initial sample is decreased compared to individual testing protocols increases the number of false negatives introduced by this second effect.
Therefore, we strongly suggest that any lab using pooled testing regularly monitor the rate of excess false negatives introduced by the pooled testing procedure.

One way to monitor the rate of introduced false negatives is as follows.  Every day, test $x\%$ of the test population individually for some small $x$, compute the false negative rate, and compare to the false negative rate in the $(100-x)\%$ that are tested in pools.
This gives a direct estimate of the number of false negatives.  However, it requires a substantial number of additional tests to compute both false negative rates.

Here we describe an indirect method of estimation, which can also be used to help dynamically set pool sizes and uses fewer total number of tests, as it does not require computation of false negative rates directly, which require verifying tests.
In other sections, we give formulas and guidelines for setting theoretically optimal pool sizes, in a setting where all tests are assumed to be perfect.
If one wishes to factor in the introduction of false negatives to the choice of pool size, the following provides a way to do so.

In this analysis, we assume that the amount of liquid taken from each sample for PCR in individual testing is $\ell$ and that in pooled testing with pools of size $n$, the amount of liquid taken from each sample for PCR is $\ell/n$.  

Every day, individually test $x\%$ of the test population.
Record the Ct values for each individual.  Convert these values into the expected concentration $c$ of viral particles in the initial sample of volume $T$, for each individual.
This part of the calculation is protocol specific, and must be developed by each lab.

To estimate the false negative rate for individual tests $f_n^I$, we
view the initial sample of volume $T$ divided into $\frac\ell T$ parts
(one of which is the part taken for PCR), and the approximately
$c \cdot T$ viral particles uniformly distributed among the parts.
This gives the estimate
\begin{equation*}
  f_n^I \approx \left(1 - \frac\ell T\right)^{c \cdot T}.
\end{equation*}
If the prevalence is $p$, we should expect
\begin{equation*}
  \frac{np}{1 - (1 - p)^n}
\end{equation*}
positive individuals per pool, and we similarly get the following
estimate for the false negative rate for pooled tests $f_n^G$:
\begin{equation*}
  f_n^G \approx \left(1 - \frac\ell{n \cdot T}\right)^{\frac{cTnp}{1 - (1 - p)^n}}
\end{equation*}

If the introduced false negative rate $f_n^G-f_n^I$ is above the desired
threshold, reduce group size, run model again, repeat until
$f_n^G-f_n^I$ is satisfactory.

\begin{figure}[H]
\label{josh_bars}
\includegraphics[width=1.0\columnwidth]{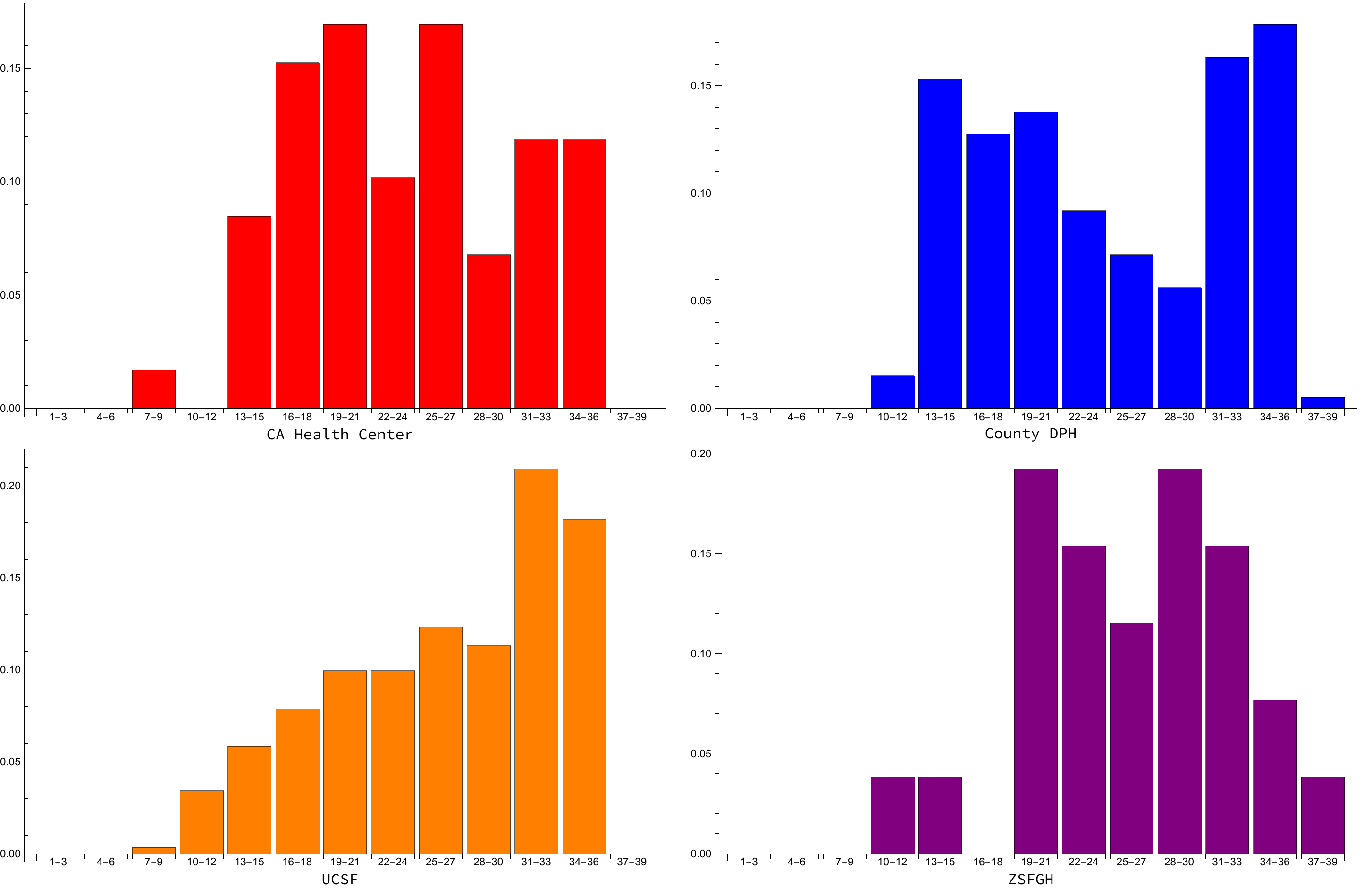}
\caption{Monitoring dilution effects is an important part of implementing group testing.
When group testing is used, Ct values should be monitored.
Here, for example, are Ct distributions from four different labs in the Bay Area in May 2020
Note that distributions of Ct values can vary substantially between labs and test populations.}
\end{figure}

% old
%\begin{figure}
%\label{josh_bars}
%\includegraphics[width=1.0\columnwidth]{josh_bars}
%\caption{Distributions of Ct values can vary substantially between labs and test populations.
%Here is an example of Ct distributions from four different labs in the Bay Area in May 2020.}
%\end{figure}

In practice, the distribution of Ct values can vary substantially between labs and between test populations.
The amount by which pooling samples increases the false negative rate depends on the distribution of Ct values in the given setting, and should be closely monitored when implementing group testing.
Figure \ref{josh_bars} shows the distribution of Ct values at four different labs in the Bay Area in May 2020.

\section{Additional architectures}
\label{s:addl_architectures}

\subsection{Early stopping}

There is a well-known modification of Dorfman testing that was suggested by Sterrett in \cite{sterrett1957}.
It is based on the assumption that when a batch turns out positive,
most likely only a small number of individuals in the batch are positive.
The method follows Steps 1--3 of Dorfman's approach identically (see
Section~\ref{ss:dorfman}), but proceeds differently with each
pooled sample:
\begin{enumerate}
\item For each pooled sample which tests negative, we deduce that all
  of the individuals in the corresponding batch are negative.

\item For each pooled sample which tests positive, we consecutively
  test the $b$ individuals in the corresponding batch.

\item If the test of the sample of the $i$-th individual is negative,
  move to the next individual.
  If the test is positive, pool the remaining $b - i$ samples, and
  repeat the process with this new pooled sample.
\end{enumerate}

While it is harder to find formulas for the optimal batch size $b$ and
the expected number of tests per hundred people, Sterrett computed
them for many useful values of the incidence rate $\rho$.
Sterrett's approach offers moderate savings over Dorfman's in the
number of tests.
On the other hand, it is a more complicated process, and takes longer
to complete the tests, because the individuals in positive batches
need to be tested sequentially.
This becomes more and more important the larger the batch size $b$.

\subsection{Double pooling}

In \cite{google}, Broder--Kumar propose \emph{double pooling}, which is a generalization of array testing.

\subsection{Compressed sensing and matrix reconstruction}

\emph{Compressed sensing} is a standard problem in the field of information theory.
The basic setup is that a sparse signal of 1's is encoded in a collection of $N$ binary digits, and we must reconstruct this signal from as few projections as possible.
Each projection adds together certain subsets of the $N$ digits.
This problem bears obvious similarities to the group testing problem, except that for group testing, addition is replaced by logical ``or''.
Techniques from compressed sensing have been applied to group testing, for instance in \cite{compressed}.

\section{Guidelines}
\label{s:guidelines}

\medskip

In this section, we consider several different prevalence regimes, and make recommendations in each regime as to which group testing architecture produces the greatest efficiency gain.
It is impossible for us to consider all possible parameter values, since there are so many parameters: true prevalence $\rho$, maximum pool size, total number of individuals $n$, etc.\
For the purposes of this subsection, we make the simplifying assumptions that the largest allowed batch size is 8.

\subsection{Classification}

In the following log-log plot, we compare the efficiency gain produced by using Dorfman testing, array testing, and hypercube testing with $d=3$ and with a batch size of 8 (in the case of Dorfman testing, the minimum of 8 and the optimal batch size).
By \emph{efficiency gain}, we mean the number of people that can be tested on average using a single test.

\begin{figure}[H]
\includegraphics[width=0.8\columnwidth]{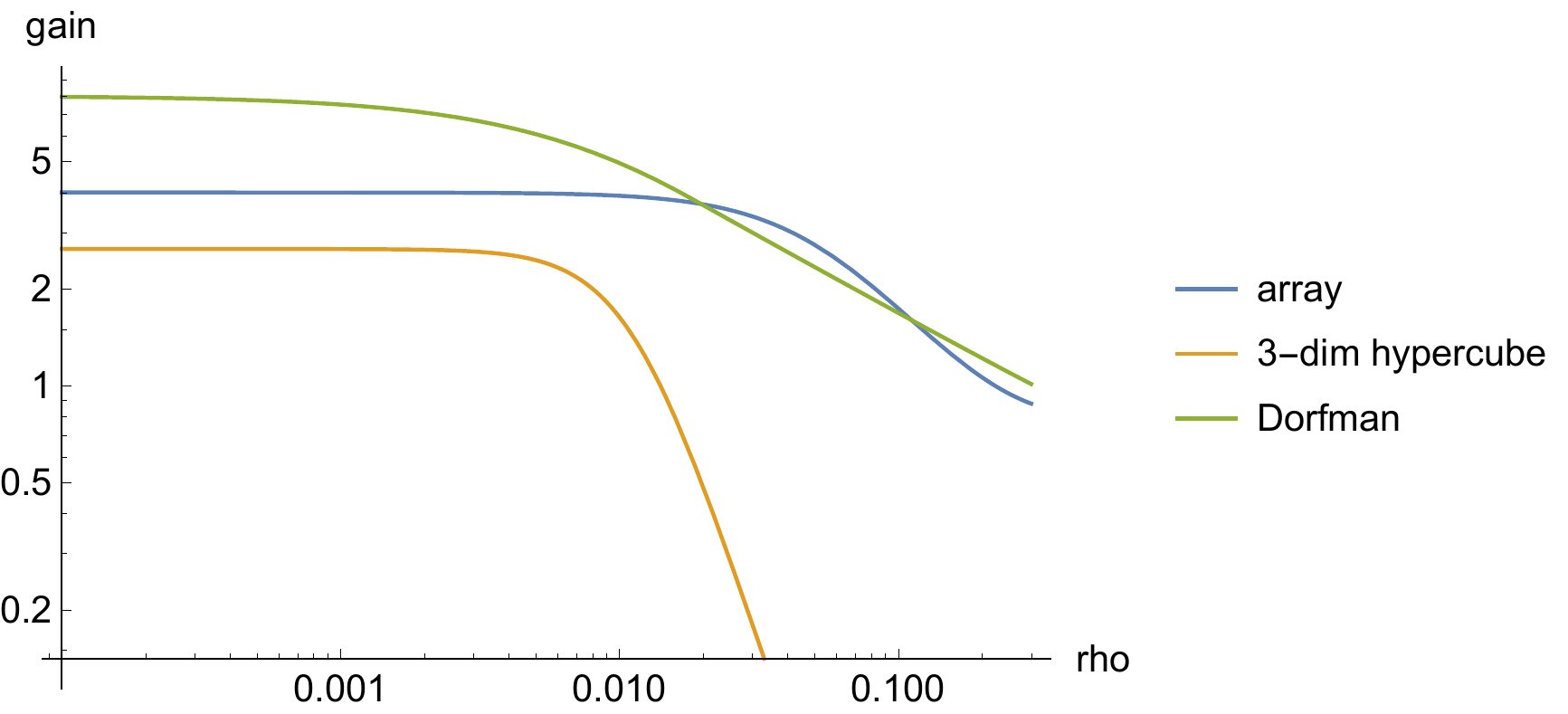}
\caption{Efficiency gains of three different architectures.
For array and 3-dimensional hypercube testing, we set the batch size at 8.
For Dorfman testing, we set it to be the minimum of 8 and the optimal batch size.}
\end{figure}

\noindent
For most choices of $\rho$, Dorfman testing produces the greatest efficiency gain.
There is a narrow range, between $\rho = 1.96\%$ and $\rho = 11.17\%$, where array testing outperforms Dorfman; however, the difference in performance in this range is minimal.

\subsection{Estimation}

We recommend using Gibbs--Gower testing for prevalence estimation.
As a rule of thumb, we suggest testing $6/p$ pools of size $8$, where $p$ is a guess for the prevalence rate.
This works well for prevalences lower than about $10\%$.
For prevalences larger than $10\%$, we suggest testing $12/p$ pools of size $4$.
The following table shows the normalized root mean square error (root mean square error divided by prevalence) when using this rule of thumb.
We see that it stays close to $15\%$ for prevalences between $0.1\%$ and $30\%$.

\medskip

\begin{center}
\begin{tabular}{|c||c|c|c|c|c|c|} \hline
  Prevalence rate & 0.1\% & 1\% & 5\% & 10\% & 10\% & 30\% \\ \hline
  Pool size & 8 & 8 & 8 & 8 & 4 & 4 \\ \hline
  Number of tests & 6000 & 600 & 120 & 60 & 120 & 40 \\ \hline
  NRMSE & 14.5\% & 14.6\% & 15.5\% & 17\% & 14.9\% & 17.3\% \\ \hline
\end{tabular}
\end{center}

\medskip

\section{Examples}
\label{s:examples}

\subsection{Examples in the context of the identification problem}

The most important features in determining which group testing architectures are best suited to a given testing scenario are:
\begin{itemize}
\item
the positive rate $\rho$ of the test population;

\item
the false positive rate $f_p$ of the test;

\item
the false negative rate $f_n$ of the test; and

\item
the time and cost to run each leg of the test.
\end{itemize}

\noindent
We begin with an overview of the current values of these parameters for the COVID-19 qPCR assay, which is a common way to test for active COVID-19 infections.
In all of the pooling schemes we will discuss below, we assume just one swab is taken from each patient, and that all the subsamples we refer to are subsamples of the liquid that is held in the lab.

\begin{itemize}
\item
For COVID-19, a wide range of test population positive rates $\rho$ are being observed around the world.
As of April 2020, $\rho$ values varied greatly from one community to another, and values from 0.1\%--10\% were commonly observed.

\smallskip

\item
The false positive rate of the COVID-19 qPCR assay is very low, and in this note we do not discuss variations on group testing that are designed to address false positives.  

\smallskip

\item
The false negative rate of the COVID-19 qPCR assay can be high --- as high as 30\%.
False negatives can occur if a patient is infected but the sampling method fails to capture any viral particles.  They can also occur when viral particles are present in the initial sample, but when a small volume of liquid is subsampled to perform qPCR, the number of viral particles in the subsample is below the limit of detection.  For individuals with low viral loads, this can be an issue and one which is exacerbated by group testing.  This, along with a method for monitoring false negatives introduced by pooled testing, is discussed in more detail in Section \ref{s:dilution}.  
\smallskip

\item
In the COVID-19 qPCR assay, a nasal swab is taken from the patient and then the sample is sent to a lab.
There it is immersed in liquid, and a sample of the liquid is plated.
Then the RNA is extracted (this might take 1--2 hours) and PCR is run (another 1--2 hours). 
\end{itemize}

Next, we do some example calculations comparing optimal tests for different variations on the Dorfman method for three hypothetical COVID-19 testing labs, with expected $\rho$ values of 3 in 10, 3 in 100, and 3 in 1000.
We do not consider variations designed to address false positives and false negatives, because for COVID-19 testing, false positive and false negative rates are low.

\medskip

\begin{center}
\begin{tabular}{|c|c|c|}
\hline
architecture & optimal batch size & individuals tested per test
\\
\hline\hline
\multicolumn{3}{|c|}{$\rho=0.3$}
\\
\hline\hline
simple Dorfman & 3 & 1.01
\\
\hline
Sterrett testing & 2 & {\bf 1.11}
\\
\hline
batched array testing & N/A & $<1$
\\
\hline\hline
\multicolumn{3}{|c|}{$\rho=0.03$}
\\
\hline\hline
simple Dorfman & 6 & 3.03
\\
\hline
Sterrett testing & 9 & 3.70
\\
\hline
batched array testing & 12 & {\bf 3.84}
\\
\hline\hline
\multicolumn{3}{|c|}{$\rho=0.003$}
\\
\hline\hline
simple Dorfman & 19 & 9.09
\\
\hline
Sterrett testing & 30 & 12.50
\\
\hline
batched array testing & 52 & {\bf 16.84}
\\
\hline
\end{tabular}
\end{center}

\bigskip

\subsection{Examples in the context of the estimation problem}

In this section, we consider several choices of prevalence rates, and compare the accuracy of the estimate produced the standard method of testing a random sample of the population to two variations.
In the first, a larger random sample of the population is tested using the same number of tests, by using Dorfman testing.
In the second, an even larger random sample of the population is tested by the same number of tests, using the Gibbs-Gower testing methodology just described.
Specifically, we make two comparisons:
\begin{itemize}

\item
In Table~\ref{t:rmse}, we compare the root mean squared error of the same three testing architectures.

\item
In Table~\ref{t:rmse_0.15}, we compare the number of tests needed to produce a root mean squared error of less than $15\%$ of the prevalence.
\end{itemize}
In both settings, we find that Gibbs--Gower testing outperforms Dorfman testing, which outperforms non-group testing.

\begin{table}[H]
\caption{Root MSE of the prevalence estimate produced by two architectures, under several different true prevalences, if we have 100 total tests at our disposal.
Again, for the two group testing architectures, we used the optimal group size.
\label{t:rmse}}
\begin{tabular}{|c|c|c|c|}
\hline
prevalence & non-group testing & Dorfman testing & Gibbs--Gower testing
\\
\hline\hline
5\% & $2.18\times 10^{-2}$ & $1.42\times10^{-2}$ & $6.28 \times 10^{-3} \:\: (b=28)$
\\
\hline
1\% & $9.95\times10^{-3}$ & $4.40\times10^{-3}$ & $1.28 \times 10^{-3} \:\: (b=143)$
\\
\hline
0.1\% & $3.16\times10^{-3}$ & $7.92\times10^{-4}$ & $1.29\times10^{-4} \:\: (b=1428)$
\\
\hline
0.01\% & $1.00\times10^{-3}$ & $1.41\times10^{-4}$ & $1.29 \times 10^{-5} \:\: (b=13726)$
\\
\hline
\end{tabular}
\end{table}

\begin{table}[H]
\caption{How many tests are needed to obtain a root MSE of $15\%$ of the prevalence.
\label{t:rmse_0.15}}
\begin{tabular}{|c|c|c|c|}
\hline
prevalence & non-group testing & \shortstack{Gibbs--Gower testing\\with group size 5} & \shortstack{Gibbs--Gower testing\\ with optimal group size}
\\
\hline\hline
5\% & 845 & 189 & 73 ($b=27$)
\\
\hline
1\% & 4400 & 899 & 76 ($b=138$)
\\
\hline
0.1\% & 44400 & 8899 & 77 ($b\approx1320$)
\\
\hline
0.01\% & 444400 & $88899$ & 79 ($b\approx12150$)
\\
\hline
\end{tabular}
\end{table}

The regimes we have considered here span a range of realistic prevalence rates for COVID-19, and we see that in all of them, Gibbs--Gower testing provides the most accurate estimation of prevalence rates given a fixed number of tests.  
At low prevalence rates, the optimal group size for Gibbs--Gower testing can be quite large --- for example, when $\rho=0.01\%$, the optimal group size is approximately 12K.
Of course, in practical settings, not only are the chemical tests limited and expensive to run, but sample collection can also be a limiting factor.  

Nevertheless, in Table~\ref{t:rmse_0.15} we see that even with a fixed group size of 5, Gibbs--Gower testing can achieve 15\% accuracy with approximately $1/5$ the number of tests required by the standard method.  
Furthermore, we can optimize the Gibbs--Gower method to minimize any function of the number of samples and the number of chemical tests.
For example, suppose in some locality the cost to run a chemical test is about ten times the cost to collect each sample.
One can optimize the Gibbs--Gower method to minimize $f(s,t) \coloneqq s+10t$, where $s=bt$ is the number of samples collected, and $t$ is the number of tests run.
We give an example calculation in Table~\ref{t:optimized}.  

\begin{table}[H]
\caption{We minimize the quantity $bt+10t$, subject to the constraint that Gibbs--Gower testing gives a root MSE of $15\%$ of the prevalence.
All numbers are approximate.
\label{t:optimized}}
\begin{tabular}{|c|c|c|c|}
\hline
prevalence & optimal group size & total tests & total samples
\\
\hline\hline
5\% & 13 & 93 & 1209
\\
\hline
1\% & 37 & 145 & 5365
\\
\hline
0.1\% & 131 & 363 & 47553
\\
\hline
\end{tabular}
\end{table}

Given the importance of daily monitoring of the prevalence rate of COVID-19 in every locality to allow early detection of local hotspots, Gibbs--Gower testing provides an important tool for regular prevalence monitoring of COVID-19.

\section{Conclusion}
\label{s:conclusion}

In this paper, we have seen that group testing has significant potential to increase COVID-19 testing capacity, given a limited number of tests.
We have restricted ourselves to qPCR testing, which is the standard method for detecting active cases of COVID-19.
We expect group testing could also be helpful in antibody testing for COVID-19, but we defer to a future paper a discussion of the concerns specific to that application.  

We believe that group testing should immediately be deployed in the US.
The current capacity in the U.S. is for testing approximately 250K people per day \cite{covid_tracking} (as of May 5, 2020).
Meanwhile, \cite{jha_tsai_jacobson} have recently suggested that the necessary testing threshold that must be met before it is possible to safely reopen the economy is the ability to test 500K people per day, representing a factor of approximately 2.
At a positive rate of 10\%, the methods we describe here can allow approximately 2 times as many people be tested with a fixed number of test kits, which would achieve this target factor of 2.

\appendix

\section{Calculations}
\label{s:formulas}

\subsection{Optimal group size for Dorfman testing}

The optimal batch size $b$ has been studied in \cite{dorfman} and \cite{williams}, among others.
Making no claims of originality, we now derive the optimal choice of $b$, supposing that the incidence rate is $\rho$.
The likelihood of a given batch containing at least one positive individual is $P \coloneqq 1 - (1-\rho)^b$, hence the expected number of positive batches will be
\begin{align}
\sum_{i=0}^{n/b}
i{{n/b} \choose i}P^i(1-P)^{n/b-i}
=
\frac nbP = \frac nb\left(1 - (1-\rho)^b\right).
\end{align}
The expected number of tests is therefore
\begin{align}
T(b)
\coloneqq
n\left(\frac 1b + \left(1 - (1-\rho)^b\right)\right).
\end{align}
The minimum is achieved at
\begin{align}
b_0
&=
\frac{2W_0\left(-\tfrac12\sqrt{-\log(1-\rho)}\right)}{\log(1-\rho)}
\\
&\approx
\frac1{\sqrt\rho} + \frac12 + \frac18\sqrt\rho + \frac13\rho + \cdots,
\nonumber
\end{align}
where $W_0$ denotes the principal branch of the Lambert W-function (i.e.\ the product logarithm), and this minimum value is
\begin{align}
T(b_0)
&=
\left(
1
-
\exp\left(2W_0\left(-\tfrac12\sqrt{-\log(1-\rho)}\right)\right)
+
\frac{\log(1-\rho)}{2W_0\left(-\tfrac12\sqrt{-\log(1-\rho)}\right)}
\right)n
\\
&\approx
\left(2\sqrt\rho - \frac12\rho + \frac5{12}\rho\sqrt\rho - \frac7{24}\rho^2+\cdots\right)n
\nonumber
\\
&\approx
2\sqrt\rho \cdot n.
\nonumber
\end{align}
Note in particular that $b_0$ depends only on $\rho$, not on $n$ (!).

We note that when there are significant rates of false positives or false negatives, the optimal batch size and expected number of tests may differ substantially from the formulas we have derived here.
This situation has been considered in \cite{hanel_thurner}.

\subsection{Array testing for the identification problem}

We now calculate the total number of tests used when array testing is applied to classify $n$ individuals, where we group those individuals in clusters of $b^2$ and where the true prevalence is $\rho$.
First, we consider the number of tests used on a single batch.
To test all rows and columns, we use $2b$ tests.
To determine the number of tests used in the next stage, we must compute the expected number of pairs $(i,j)$ such that the $i$-th row and the $j$-column both test positive.
We can approximate this by assuming that the number of positive rows and the number of positive columns are independent:
\begin{align}
\left(\left(1-(1-\rho)^b\right)b\right)^2
=
\left(1-(1-\rho)^b\right)^2b^2.
\end{align}
It follows that an approximation for the expected total number of tests that we must use on this single batch is
\begin{align}
2b + \left(1-(1-\rho)^b\right)^2b^2.
\end{align}
The approximate expected total number of tests we use on all $n/b^2$ clusters is therefore
\begin{align}
\label{eq:num_tests_unbatched_array}
\frac{2n}b + \left(1-(1-\rho)^b\right)^2n
=
\left(\frac2b + \left(1 - (1-\rho)^b\right)^2\right)n.
\end{align}

\subsection{Hypercube testing for the identification problem}

Next, we perform a similar calculation as in the previous subsection: we compute the total number of tests used when we apply $d$-dimensional hypercube testing to classify $n$ individuals, where we group those individuals in clusters of $b^d$.
To test all subsets of the form $\bigl\{(i_1,\ldots,i_{j-1},?,i_{j+1},\ldots,i_d)\bigr\}$ (call these ``$j$-rows'' from now on), we use $db^{d-1}$ tests.
Making a similar independence approximation as above, we see that the expected number of individuals $(i_1,\ldots,i_d)$ such that the $(i_1,\ldots,i_{j-1},i_{j+1},\ldots,i_d)$-th $j$-row tests positive for each $j$ is
\begin{align}
\left(\left(1-(1-\rho)^b\right)b^{d-1}\right)^d
=
\left(1-(1-\rho)^b\right)^db^{d(d-1)}.
\end{align}
An approximation for the expected total number of tests we must use on a single cluster is therefore
\begin{align}
db^{d-1} + \left(1-(1-\rho)^b\right)^db^{d(d-1)},
\end{align}
so the approximate expected total number of tests we use on all $n/b^d$ clusters is
\begin{align}
\left(\frac db + \left(1 - (1-\rho)^b\right)^db^{d(d-2)}\right)n.
\end{align}
Note in particular that if $b = 8$ and $\rho$ is very small, this last quantity is approximately
\begin{align}
\frac d8\cdot n
\end{align}

\subsection{The MSE for estimation with the Gibbs--Gower technique}

The estimate $\hat p$ is biased, and more precisely an overestimate (when $b > 1$) with expected value
\begin{equation*}
E(\hat p)
=
1 - \sum_{i = 0}^t \left(\frac it \right)^{1/b} {t\choose i} \bigl((1 - p)^b\bigr)^i \bigl(1 - (1 - p)^b\bigr)^{t - i}.
\end{equation*}
The mean squared error (MSE) of the estimate is given by
\begin{equation*}
  E((\hat p - p)^2)
  =
  \sum_{i = 0}^t
  \left(\frac it \right)^{2/b} {t \choose i} ((1 - p)^b)^i (1 - (1 - p)^b)^{t - i}
  +
  (p - 1)\bigl(p + 1 - 2E(\wh p)\bigr).
\end{equation*}
An approximation of this is the asymptotic variance
\begin{equation*}
\frac{1 - (1 - p)^b}{t b^2 (1 - p)^{b - 2}}.
\end{equation*}

When $b = 1$, the expected value is $E(\hat p) = p$, and the MSE specializes to
\begin{equation*}
E((\hat p - p)^2)
=\frac{p(1 - p)}t.
\end{equation*}

\end{document}